\documentclass[journal]{IEEEtran}

\usepackage{epsfig}
\usepackage[tight,footnotesize]{subfigure}

\usepackage[cmex10]{amsmath}
\newtheorem{theorem}{Theorem}
\def \endproof { \hfill \rule{1,5truemm}{2truemm} \smallskip}
\def \begproof {\vspace{0truecm} \noindent {\sc Proof}: }

\begin{document}

\title{An Information-Theoretical View of Network-Aware Malware Attacks}

\author{Zesheng~Chen,~\IEEEmembership{Member,~IEEE,}
        and~Chuanyi~Ji,~\IEEEmembership{Senior~Member,~IEEE}
\thanks{ Part of the material in this paper was presented in
\cite{metric} at the IEEE INFOCOM'07 conference.

Z. Chen is with the Department of Electrical and Computer
Engineering, Florida International University, Miami, FL, 33174 USA
e-mail: zchen@fiu.edu. C. Ji is with the Department of Electrical
and Computer Engineering, Georgia Institute of Technology, Atlanta,
GA, 30332 USA e-mail: jic@ece.gatech.edu.}}

\maketitle

\begin{abstract}
This work investigates three aspects: (a) a {\it network
vulnerability} as the non-uniform vulnerable-host distribution, (b)
{\it threats}, {\em i.e.}, intelligent malwares that exploit such a
vulnerability, and (c) {\it defense}, {\em i.e.}, challenges for
fighting the threats. We first study five large data sets and
observe consistent clustered vulnerable-host distributions. We then
present a new metric, referred to as the {\it non-uniformity
factor}, which quantifies the unevenness of a vulnerable-host
distribution. This metric is essentially the Renyi information
entropy and better characterizes the non-uniformity of a
distribution than the Shannon entropy. Next, we analyze the
propagation speed of network-aware malwares in view of information
theory. In particular, we draw a relationship between Renyi
entropies and randomized epidemic malware-scanning algorithms. We
find that the infection rates of malware-scanning methods are
characterized by the Renyi entropies that relate to the information
bits in a non-unform vulnerable-host distribution extracted by a
randomized scanning algorithm. Meanwhile, we show that a
representative network-aware malware can increase the spreading
speed by exactly or nearly a non-uniformity factor when compared to
a random-scanning malware at an early stage of malware propagation.
This quantifies that how much more rapidly the Internet can be
infected at the early stage when a malware exploits an uneven
vulnerable-host distribution as a network-wide vulnerability.
Furthermore, we analyze the effectiveness of defense strategies on
the spread of network-aware malwares. Our results demonstrate that
counteracting network-aware malwares is a significant challenge for
the strategies that include host-based defense and IPv6.
\end{abstract}


\section{Introduction}
\label{sec:introduction}

{\it Malware} attacks are a significant threat to networks. Malwares
are malicious softwares that include worms, viruses, bots, and
spywares. A fundamental characteristic of malwares is
self-propagation, {\em i.e.}, a malware can infect vulnerable hosts
and use infected hosts to self-disseminate. A key component of
malware propagation is malware-scanning methods, {\em i.e.}, how
effectively and efficiently the malware finds vulnerable targets.
Most of the real, especially old worms, such as Code Red
\cite{Moore:02}, Slammer \cite{Moore:03}, and latter Witty
\cite{Shannon:04}, exploit naive random scanning. Random scanning
chooses target IP addresses uniformly and does not take any
information on network structures into consideration. Advanced
scanning methods, however, have been developed that exploit the IP
address structure. For example, Code Red II and Nimda worms have
used localized scanning \cite{CERT_codered,CERT_nimda}. Localized
scanning preferentially searches for vulnerable hosts in the local
sub-network. The Blaster worm has used sequential scanning in
addition to localized scanning \cite{Blaster}. Sequential scanning
searches for vulnerable hosts through their closeness in the IP
address space. The AgoBot has employed a blacklist of the well-known
monitored IP address space and avoided scanning these addresses to
be stealthy \cite{Rajab:06}. A common characteristic of these
malwares is that they scan for vulnerable hosts by taking a certain
structure in the IP address space into consideration. Such a
structure, as we shall soon show, exhibits network vulnerabilities
to defenders and advantages to attackers.

In this paper, we study the perspective of attackers who attempt to
collect the information on network vulnerabilities and design
intelligent malwares. By studying this perspective, we hope to help
defenders better understand and defend against malware propagation.
For attackers, an open question is how certain information can help
them design fast-spreading malwares. The information may include the
vulnerability on end hosts, the number of vulnerable hosts, the
locations of detection systems, and the distributions of vulnerable
hosts.

This work focuses on vulnerable-host distributions. The
vulnerable-host distributions have been observed to be bursty and
spatially inhomogeneous by Barford {\em et al.} \cite{Paul:06}. A
non-uniform distribution of Witty-worm victims has been reported by
Rajab {\em et al.} \cite{Rajab}. A Web-server distribution has been
found to be non-uniform in the IP address space in our prior work
\cite{IS1}. These discoveries suggest that vulnerable hosts and Web
servers may be ``clustered" ({\em i.e.}, non-uniform). The
clustering/non-uniformity makes the network vulnerable since if one
host is compromised in a cluster, the rest may be compromised rather
quickly. Therefore, the information on vulnerable-host distributions
can be critical for attackers to develop intelligent malwares.

We refer the malwares that exploit the information on highly uneven
distributions of vulnerable hosts as {\it network-aware} malwares.
Such malwares include aforementioned localized-scanning and
sequential-scanning malwares. In our prior work, we have studied
{\it importance-scanning} malwares \cite{IS1,IS2,Gu:07}.
Specifically, importance scanning provides a ``what-if" scenario:
When there are many ways for network-aware malwares to exploit the
information on vulnerable hosts, importance scanning is a worst-case
threat-model and can serve as a benchmark for studying real
network-aware malwares. What has been observed is that real
network-aware and importance-scanning malwares spread much faster
than random-scanning malwares \cite{Rajab,IS1}. This shows the
importance of the problem. It is not well understood, however, how
to characterize the relationship between the information on
vulnerable-host distributions and the propagation speed of
network-aware malwares.

Questions arise. Does there exist a {\it generic} characteristic
across different vulnerable-host distributions? If so, how do
network-aware malwares exploit such a vulnerability? How can we
defend against such malwares? Our goal is to investigate such a
generic characteristic in vulnerable-host distributions, to quantify
its relationship with network-aware malwares, and to understand the
effectiveness of defense strategies. To achieve this goal, we
investigate network-aware malware attacks in view of information
theory, focusing on both the worst-case and real network-aware
malwares.

A fundamental concept of information theory is the {\it entropy}
that measures the uncertainty of outcomes of a random event. The
reduction of uncertainty is measured by the amount of acquired
information. We apply the {\it Renyi entropy}, a generalized entropy
\cite{Renyi}, to analyze the uncertainty of finding vulnerable hosts
for different malware-scanning methods. This would relate
malware-attacking methods with the information bits extracted by
malwares from the vulnerable-host distribution.

As the first step, we study, from five large-scale measurement sets,
the common characteristics of non-uniform vulnerable-host
distributions. Then, we derive a new metric as the {\it
non-uniformity factor} to characterize the non-uniformity of a
vulnerable-host distribution. A larger non-uniformity factor
reflects a more non-uniform distribution of vulnerable hosts. We
obtain the non-uniformity factors from the data sets on
vulnerable-host distributions and show that all data sets have large
non-uniformity factors. Moreover, the non-uniformity factor is a
function of the Renyi entropies of order two and zero \cite{Renyi}.
We show that the non-uniformity factor better characterizes the
unevenness of a distribution than the Shannon entropy. Therefore, in
view of information theory, the non-uniformity factor provides a
quantitative measure of the unevenness/uncertainty of a
vulnerable-host distribution.

Next, we relate the generalized entropy with network-aware scanning
methods. The class of network-aware malwares that we study all
utilizes {\em randomized} epidemic algorithms. Hence the importance
of applying the generalized entropy is that the Renyi entropy
characterizes the bits of information extractable by the randomized
epidemic algorithms. Specifically, we explicitly relate the Renyi
entropy with the randomized epidemic scanning methods through
analyzing the spreading speed of network-aware malwares at an early
stage of propagation. A malware that spreads faster at the early
stage can in general infect most of the vulnerable hosts in a
shorter time. The propagation ability of a malware at the early
stage is characterized by the {\it infection rate} \cite{Zou}. We
derive the infection rates of a class of network-aware malwares. We
find that the infection rates of random-scanning and network-aware
malwares are determined by the uncertainty of the vulnerable-host
distribution or the Renyi entropies of different orders.
Specifically, a random-scanning malware has the largest uncertainty
({\em e.g.}, Renyi entropy of order zero), and an optimal
importance-scanning malware has the smallest uncertainty ({\em
e.g.}, Renyi entropy with order infinity). Moreover, the infection
rates of some real network-aware malwares depend on the
non-uniformity factors or the Renyi entropy of order two. For
example, compared with random scanning, localized scanning can
increase the infection rate by nearly a non-uniformity factor.
Therefore, the infection rates of malware-scanning algorithms are
characterized by the Renyi entropies, relating the efficiency of a
randomized scanning algorithm with the uncertainty on a non-uniform
vulnerable-host distribution. These analytical results on the
relationships between vulnerable-host distributions and
network-aware malware spreading ability are validated by simulation.

Finally, we study new challenges to malware defense posed by
network-aware malwares. Using the non-uniformity factor, we show
quantitatively that the host-based defense strategies, such as
proactive protection \cite{Brumley} and virus throttling
\cite{Twycross}, should be deployed at almost all hosts to slow down
network-aware malwares at the early stage. A partial deployment
would nearly invalidate such host-based defense. Moreover, we
demonstrate that the infection rate of a network-aware malware in
the IPv6 Internet can be comparable to that of the Code Red v2 worm
in the IPv4 Internet. Therefore, fighting network-aware malwares is
a real challenge.

The remainder of this paper is structured as follows. Section
\ref{sec:prelim} introduces information theory and malware
propagation. Section \ref{sec:data} presents our collected data
sets. Section \ref{sec:formulation} formulates the problems studied
in this work. Section \ref{sec:factor} introduces a new metric
called the non-uniformity factor and compares this metric to the
Shannon entropy. Sections \ref{sec:ability} and
\ref{sec:simulations} characterize the spreading ability of
network-aware malwares through theoretical analysis and simulation.
Section \ref{sec:applications} further studies the effectiveness of
defense strategies on network-aware malwares. Section
\ref{sec:conclusions} concludes this paper.

\section{Preliminaries}
\label{sec:prelim}

In this section, we provide the background on information theory and
malware propagation.

\subsection{Renyi Entropy and Information Theory}

An entropy is a measure of the average information uncertainty of a
random variable \cite{info}. A general entropy, called the Renyi
entropy \cite{Renyi,Cachin}, is defined as
\begin{equation}
  H_q(X) = \frac{1}{1-q}\log_2{\sum_{x\in \mathcal{X}}\left( P_X(x)
  \right)^q},\ \mbox{for}\ q \not= 1,
\end{equation}
where the random variable $X$ is with probability distribution $P_X$
and alphabet $\mathcal{X}$. The well-known Shannon entropy is a
special case of the Renyi entropy, {\em i.e.},
\begin{equation}
  H(X) = \lim_{q\rightarrow 1}{H_q(X)}.
\end{equation}
It is noted that
\begin{equation}
 H_0(X) = \log_2{|\mathcal{X}|}
\end{equation}
where $|\mathcal{X}|$ is the alphabet size, and
\begin{equation}
  H_\infty (X) = -\log_2{\max_{x\in \mathcal{X}}P_X(x)}
\end{equation}
where $H_\infty(X)$ is a result from $\lim_{q\rightarrow
\infty}{H_q(X)}$ and is called the min-entropy of $X$. In this
paper, moreover, we are also interested in the Renyi entropy of
order two, {\em i.e.},
\begin{equation}
  H_2(X) = -\log_2{\sum_{x\in \mathcal{X}}{\left( P_X(x) \right)^2}}.
\end{equation}
Comparing $H_0(X)$, $H(X)$, $H_2(X)$, and $H_\infty (X)$, we have
the following theorem that has been proved in
\cite{Renyi_proof,Cachin}.
\begin{theorem}
\label{thm:comp}
\begin{equation}
\label{ineq:renyi}
   H_0(X) \ge H(X) \ge H_2(X) \ge H_\infty (X)
\end{equation}
with equality if and only if $X$ is uniformly distributed over
$\mathcal{X}$.
\end{theorem}

Information theory has been applied in a wide range of fields, such
as communication theory, Kolmogorov complexity, and cryptography. A
fundamental result of information theory is that data compression
can be achieved by assigning short codewords to the most frequent
outcomes of the data source and necessarily longer codewords to the
less frequent outcomes \cite{info}.

\subsection{Malware Propagation}

Similar to data compression, a smart malware that searches for
vulnerable hosts can assign more scans to a range of IP addresses
that contain more vulnerable hosts. Thus, the malware can reduce the
number of scans for attacking a large number of vulnerable hosts. We
call such a malware as a {\it network-aware} malware. In essence,
network-aware malwares consider the network structure ({\em i.e.},
an uneven distribution of vulnerable hosts) to speed up the
propagation.

Many network-aware malwares have been studied. For example,
routable-scanning malwares select targets only in the routable
address space, using the information provided by the BGP routing
table \cite{Wu,Zou}. Evasive worms exploit lightweight sampling to
obtain the knowledge of {\it live} subnets of the address space and
spread only in these networks \cite{Rajab:06}. In our prior work, we
have studied a class of worst-case malwares, called {\it
importance-scanning} malwares, which exploit non-uniform
vulnerable-host distributions in an optimal fashion \cite{IS1,IS2}.
Importance scanning is developed from and named after importance
sampling in statistics. Importance scanning probes the Internet
according to an underlying vulnerable-host distribution. Such a
scanning method forces malware scans on the most relevant parts of
an address space and supplies an optimal scanning strategy.
Furthermore, if the complete information of vulnerable hosts is
known, an importance-scanning malware can achieve the top speed of
infection and become flash worms \cite{Staniford:04}.

\section{Measurements and Vulnerable-Host Distributions}
\label{sec:data}

We begin our study by considering how significant the unevenness of
vulnerable-host distributions is. We use five large data sets to
obtain empirical vulnerable-host distributions.

\subsection{Measurements}

{\it DShield (D1)}: DShield collects intrusion detection system
(IDS) logs \cite{DShield}. Specifically, DShield provides the
information of vulnerable hosts by aggregating logs from more than
1,600 IDSes distributed throughout the Internet. We further focus on
the following ports that were attacked by worms: 80 (HTTP), 135
(DCE/RPC), 445 (NetBIOS/SMB), 1023 (FTP servers and the remote shell
attacked by W32.Sasser.E.Worm), and 6129 (DameWare).

{\it iSinks (P1 and C1)}: Two unused address space monitors run the
{\it iSink} system \cite{iSink}. The monitors record the unwanted
traffic arriving at the unused address spaces that include a Class A
network (referred to as ``Provider" or P1) and two Class B networks
at the campus of the University of Wisconsin (referred to as
``Campus" or C1) \cite{Paul:06}.

{\it Witty-worm victims (W1)}: A list of Witty-worm victims is
provided by CAIDA \cite{Shannon:04}. CAIDA used a network telescope
with approximate $2^{24}$ IP addresses to log the traffic of
Witty-worm victims that are Internet security systems (ISS)
products.

{\it Web-server list (W2)}: The IP addresses of Web servers were
collected through UROULETTE (http://www.uroulette.com/). UROULETTE
provides a random uniform resource locator (URL) generator to obtain
a list of IP addresses of Web servers.

The first three data sets (D1, P1, and C1) were collected over a
seven-day period from 12/10/2004 to 12/16/2004 and have been studied
in \cite{Paul:06} to demonstrate the bursty and spatially
inhomogeneous distribution of malicious source IP addresses. The
last two data sets (W1 and W2) have been used in our prior work
\cite{IS1} to show the virulence of importance-scanning malwares.
The summary of our data sets is given in Table \ref{tab:data}.
\begin{table}[htb]
\begin{center}
\caption{Summary of the data sets.} \label{tab:data}
\begin{tabular}{|r|l|c|} \hline
   Trace       &  Description  & Number of unique source addresses       \\ \hline
   D1          &  DShield      & 7,694,291 \\
   P1          &  Provider     & 2,355,150 \\
   C1          &  Campus       & 448,894   \\
   W1          &  Witty-worm victims & 55,909 \\
   W2          &  Web servers   & 13,866       \\ \hline
\end{tabular}
\end{center}
\end{table}

\subsection{Vulnerable-Host Distributions}
\label{sec:vul_dist}

To obtain vulnerable-host group distributions, we use the classless
inter-domain routing (CIDR) notation \cite{Kohler}. The Internet is
partitioned into subnets according to the first $l$ bits of IP
addresses, {\em i.e.}, /$l$ prefixes or /$l$ subnets. In this
division, there are $2^{l}$ subnets, and each subnet contains
$2^{32-l}$ addresses, where $l\in \{0,1,\cdots,32\}$. For example,
when $l=8$, the Internet is grouped into Class A subnets ({\em
i.e.}, /8 subnets); when $l=16$, the Internet is partitioned into
Class B subnets ({\em i.e.}, /16 subnets).

\begin{figure}[bt!]
\centering \subfigure[Population distributions in /8 subnets.] {
\label{fig:ccdf_8} \psfig{figure=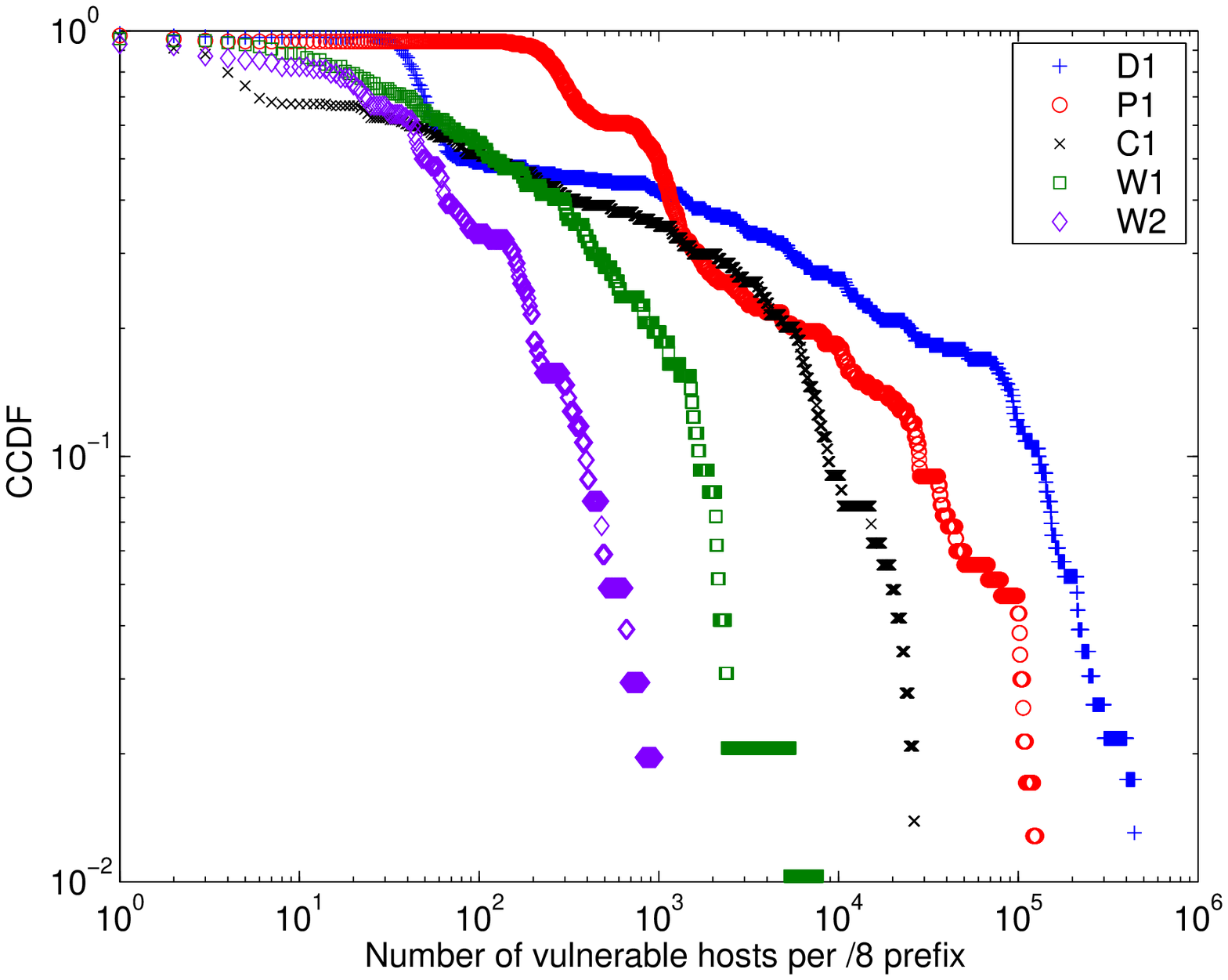,width=7.8cm} }
\subfigure[Population distributions in /16 subnets.] {
\label{fig:ccdf_16}
\psfig{figure=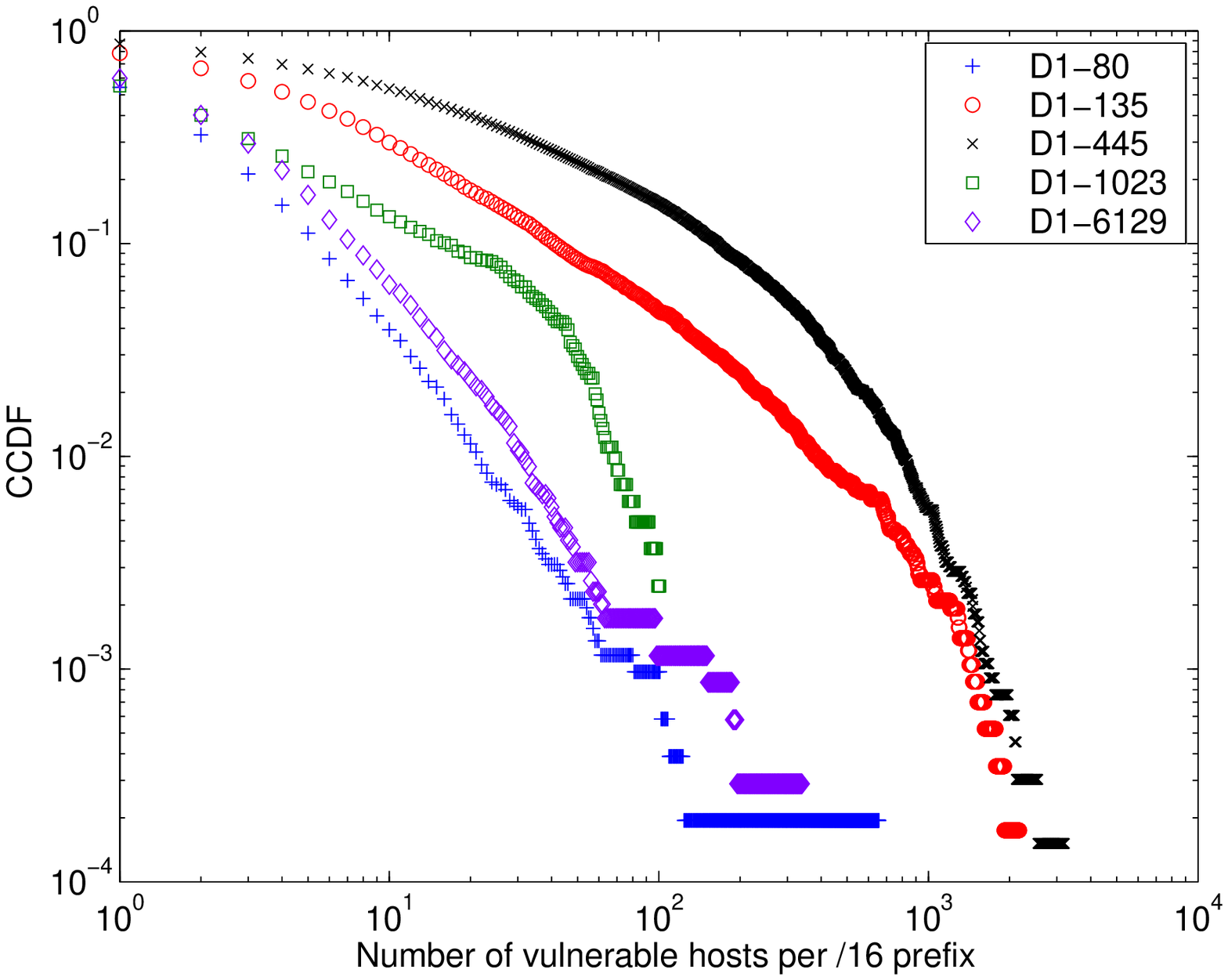,width=7.8cm} } \caption{CCDF
of collected data sets.} \label{fig:ccdf}
\end{figure}

We plot the complementary cumulative distribution functions (CCDF)
of our collected data sets in /8 and /16 subnets in Figure
\ref{fig:ccdf} in log-log scales. CCDF is defined as the faction of
the subnets with the number of hosts greater than a given value.
Figure \ref{fig:ccdf_8} shows population distributions in /8 subnets
for D1, P1, C1, W1, and W2, whereas Figure \ref{fig:ccdf_16}
exhibits host distributions in /16 subnets for D1 with different
ports (80, 135, 445, 1023, and 6129). Figure \ref{fig:ccdf}
demonstrates a wide range of populations, indicating highly
inhomogeneous address structures. Specifically, the relatively
straight lines, such as W2 and D1-135, imply that vulnerable hosts
follow a power law distribution. Similar observations were given in
\cite{Paul:06,Rajab,Pryadkin,Moore:02,Moore:03,IS1,Vojnovic}.

Why is the vulnerable-host distribution highly non-uniform in the
IPv4 address space? An answer to this question would involve other
empirical study beyond the scope of this work. Nevertheless, we
hypothesize several possible reasons. First, no vulnerable hosts can
exist in reserved or multicast address ranges \cite{IANA}. Second,
different subnet administrators make different use of their own IP
address space. Third, a subnet intends to have many computers with
the same operating systems and applications for easy management
\cite{Staniford:02,AAWP}. Last, some subnets are more protected than
others \cite{Paul:06,Rajab}.

\section{Problem Formulation}
\label{sec:formulation}

Motivated by the empirical study, we provide a problem formulation
in this section.

\subsection{Characterization}

Let a vulnerable host be a host that can be infected by a malware. A
vulnerable host can be either already infected or uninfected. In
this work, we denote vulnerable hosts as uninfected vulnerable
hosts.

We consider aggregated vulnerable-host distributions. Let $l$ ($0\le
l \le 32$) be an aggregation level of IP addresses as defined in
Section \ref{sec:vul_dist}. For a given $l$, let $N_i^{(l)}$ be the
number of vulnerable hosts in /$l$ subnet $i$, where $1 \le i \le
2^l$. Let $N$ be the total number of vulnerable hosts, where
$N=\sum_{i=1}^{2^l}{N_i^{(l)}}$. Let $p_g^{(l)}(i)$
($i=1,2,\cdots,2^l$) be the probability that a randomly chosen
vulnerable host is in the $i$-th $/l$ subnet. Then
$p_g^{(l)}(i)=\frac{N_i^{(l)}}{N}$; and
$\sum_{i=1}^{2^l}{p_g^{(l)}(i)}=1$. Thus, $p_g^{(l)}(i)$'s denote
the group distribution of vulnerable hosts in /$l$ subnets.

Now consider a malware or an adversary that searches for vulnerable
hosts. An adversary often does not have the complete knowledge on
the locations of vulnerable hosts. Hence malwares make a random
guess on which $/l$ subnets are likely to have most vulnerable
hosts. This results in a class of randomized epidemic algorithms for
malwares to scan subnets. Let $q_g^{(l)}(i)$ ($i=1,2,\cdots,2^l$) be
the probability that a malware scans the $i$-th $/l$ subnet. As we
shall see, $q_g^{(l)}(i)$'s characterize how effectively a malware
scans and thus hits vulnerable hosts.

\subsection{Examples}

Consider $/8$ subnets ($l=8$). As an extreme case if all vulnerable
hosts are in subnet $123.0.0.0/8$, $p_g^{(8)}(123)=1$ and
$p_g^{(8)}(i)=0$ for $i \ne 123$. Hence, in view of a network
defender, the network would be extremely vulnerable since if the
malware discovers this subnet, the malware could focus its scan
accordingly and potentially find all vulnerable hosts. As another
extreme case, if $p_g^{(8)}(i)={1 \over 2^8}$ is uniform, in view of
a network defender, it would be harder for the malware to find
vulnerable hosts rapidly as they are uniformly distributed over all
/8 subnets.

In view of an adversary, whether and how vulnerable hosts can be
discovered and scanned depends on a randomized algorithm utilized by
the malware. In other words, an adversary can customize
$q_g^{(l)}(i)$'s to make the malware spread either faster or slower.
Specifically, for the first extreme case where all vulnerable
subnets are concentrated in subnet $123.0.0.0/8$,
$q_g^{(8)}(i)=p_g^{(8)}(i)$ would be the best choice, and the
resulting malware would spread the fastest. But if the adversary
makes a poor choice of $q_g^{(l)}(i)$'s being uniform across /8
subnets, the resulting malware spread would be slow. Hence
$p_g^{(l)}(i)$'s and $q_g^{(l)}(i)$'s characterize the severity of
network vulnerabilities and the virulence of randomized epidemic
scanning algorithms, respectively.

\subsection{Problems}

We now consider network-aware malware attacks in view of information
theory. If a malware obtains the partial information on vulnerable
hosts ({\em e.g.}, $p_g^{(l)}(i)$'s), it can extract the information
(bits) to design randomize epidemic algorithms ({\em e.g.},
$q_g^{(l)}(i)$'s). A fundamental question is how we can relate the
information on vulnerable-host distributions with the performance of
randomized epidemic algorithms. Specifically, we intend to study the
following questions:
\begin{itemize}
\item
What information-theoretical measure can be used to quantify the
unevenness of vulnerable-host distributions?
\item
How much information (bits) on vulnerable hosts can be extracted by
randomized epidemic algorithms utilized by malwares?
\item
How good are practical randomized epidemic algorithms?
\item
What is the effectiveness of defense methods on network-aware
malwares?
\end{itemize}

\section{Non-Uniformity Factor}
\label{sec:factor}

In this section, we derive a simple metric, called the {\it
non-uniformity factor}, to quantify the vulnerability, {\em i.e.},
the non-uniformity of a vulnerable-host distribution. We show that
the non-uniformity factor is a function of Renyi entropies. We then
compare the non-uniformity factor with the well-known Shannon
entropy.

\subsection{Definition and Property}

{\bf Definition}: The {\it non-uniformity factor} in /$l$ subnets is
defined as
\begin{equation}
\label{equ:uniformity}
  \beta^{(l)}=2^l \sum_{i=1}^{2^l}{\left(p_g^{(l)}(i)\right)^2}.
\end{equation}

One property of $\beta^{(l)}$ is that
\begin{equation}
  \beta^{(l)} \ge \left(\sum_{i=1}^{2^l}{p_g^{(l)}(i)} \right)^2  =  1.
\end{equation}
The above inequality is derived by the Cauchy-Schwarz inequality.
The equality holds if and only if $p_g^{(l)}(i) = 2^{-l}$, for
$i=1,2,\cdots,2^l$. In other words, when the vulnerable-host
distribution is uniform, $\beta^{(l)}$ achieves the minimum value 1.
On the other hand, since $p_g^{(l)}(i) \ge 0$,
\begin{equation}
  \beta^{(l)} \le 2^l \cdot \left(\sum_{i=1}^{2^l}{p_g^{(l)}(i)}
  \right)^2 = 2^l.
\end{equation}
The equality holds when $p_g^{(l)}(j)=1$ for some $j$ and
$p_g^{(l)}(i) = 0$, $i\not= j$, {\em i.e.}, all vulnerable hosts
concentrate on one subnet. This means that when the vulnerable-host
distribution is extremely non-uniform, $\beta^{(l)}$ obtains the
maximum value $2^l$. Moreover, assuming that vulnerable hosts are
uniformly distributed in the first $n$ ($1 \le n \le 2^l$) /$l$
subnets, {\em i.e.}, $p_g^{(l)}(i)=\frac{1}{n}$, $i=1,2,\cdots,n$;
and $p_g^{(l)}(i)=0$, $i=n+1,\cdots,2^l$, we have
$\beta^{(l)}=\frac{2^l}{n}$. Therefore, $\beta^{(l)}$ characterizes
the non-uniformity of a vulnerable-host distribution. A larger
non-uniformity factor reflects a more non-uniform distribution of
vulnerable hosts.

The non-uniformity factor is indeed related to a distance between a
vulnerable-host distribution and a uniform distribution. Consider
$L_2$ distance between $p_g^{(l)}(i)$ and the uniform distribution $
p_u^{(l)}(i)={1\over {2^l}}$ for $i=1,2,\cdots,2^l$, where
\begin{equation}
{|| p_g^{(l)}-
p_u^{(l)}||_2}^2=\sum_{i=1}^{2^l}{\left(p_g^{(l)}(i)-\frac{1}{2^l}\right)^2},
\label{equ:uniformity-l2}
\end{equation}
which leads to
\begin{equation}
\beta^{(l)} = 2^l\cdot {||p_g^{(l)}-p_u^{(l)}||_2}^2+1.
\end{equation}
For a given $l$, $2^l$ is a constant that is the size of the sample
space of /$l$ subnets. Hence $\beta^{(l)}$ essentially measures the
deviation of a vulnerable-host group distribution from a uniform
distribution for /$l$ subnets.

How does $\beta^{(l)}$ vary with $l$? When $l=0$, $\beta^{(0)}=1$.
In the other extreme where $l=32$,
\begin{equation}
\label{equ:l_32}
 p_g^{(32)}(i) = \left\{ \begin{array}{ll}
         \frac{1}{N}, & \mbox{address $i$ is vulnerable to the malware}; \\
         0, & \mbox{otherwise}, \end{array} \right.
\end{equation}
which results in $\beta^{(32)}=\frac{2^{32}}{N}$. More importantly,
the ratio of $\beta^{(l)}$ to $\beta^{(l-1)}$ lies between 1 and 2,
as shown below.
\begin{theorem}
\label{thm:uniformity}
\begin{equation}
  \beta^{(l-1)} \le \beta^{(l)} \le 2\beta^{(l-1)},
\end{equation}
where $l \in \{1,\cdots,32\}$.
\end{theorem}

The proof of Theorem \ref{thm:uniformity} is given in Appendix 1. An
intuitive explanation of this theorem is as follows. For /$l$ and
/$(l-1)$ subnets, group $i$ ($i=1,2,\cdots , 2^{l-1}$) of /$(l-1)$
subnets is partitioned into groups $2i-1$ and $2i$ of /$l$ subnets.
If vulnerable hosts in each group of /$(l-1)$ subnets are equally
divided into the groups of /$l$ subnets ({\em i.e.},
$p_g^{(l)}(2i-1)=p_g^{(l)}(2i)=\frac{1}{2}p_g^{(l-1)}(i)$, $\forall
i$), then $\beta^{(l)}=\beta^{(l-1)}$. If the division of vulnerable
hosts is extremely uneven for all groups ({\em i.e.},
$p_g^{(l)}(2i-1)=0$ or $p_g^{(l)}(2i)=0$, $\forall i$), then
$\beta^{(l)}=2\beta^{(l-1)}$. Excluding these two extreme cases,
$\beta^{(l-1)} < \beta^{(l)} < 2\beta^{(l-1)}$. Therefore,
$\beta^{(l)}$ is a non-decreasing function of $l$. Moreover, the
ratio of $\beta^{(l)}$ to $\beta^{(l-1)}$ reflects how unevenly
vulnerable hosts in each /$(l-1)$ subnet distribute between two
groups of /$l$ subnets. This ratio is indicated by the slope of
$\beta^{(l)}$ in a $\beta^{(l)}\sim l$ figure.

\subsection{Estimated Non-Uniformity Factor}

\begin{figure}[bt!]
\centering \subfigure[Five data sets.] { \label{fig:rate_all}
\psfig{figure=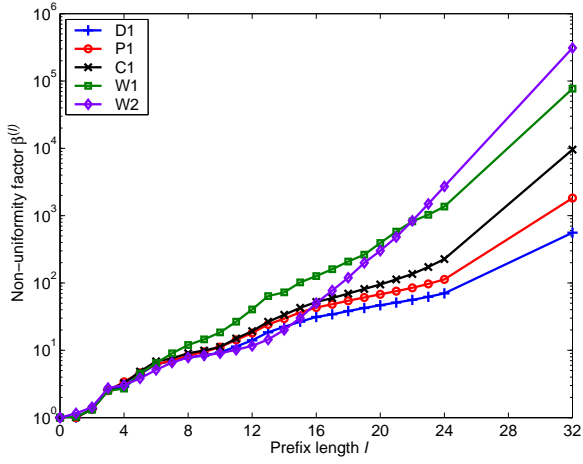,width=7.8cm} } \subfigure[D1 with
different ports.] { \label{fig:rate_dshield}
\psfig{figure=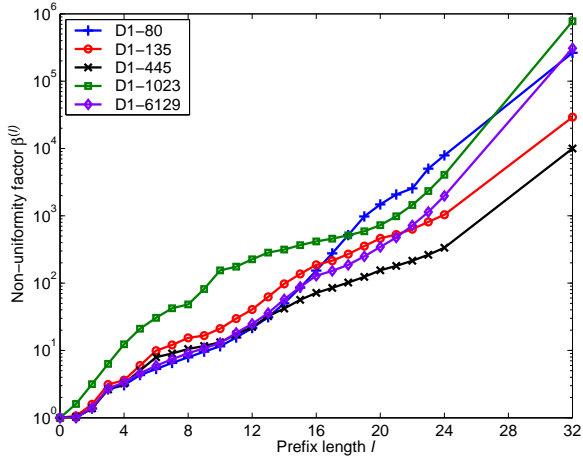,width=7.8cm} }
\caption{Non-uniformity factors of collected data sets. Note that
the y-axis uses a {\it log} scale.} \label{fig:rate}
\end{figure}

Figure \ref{fig:rate} shows the non-uniformity factors estimated
from our data sets. The non-uniformity factors increase with the
prefix length for all data sets. Note that the y-axis is in a {\it
log} scale. Thus, $\beta^{(l)}$ increases {\it almost exponentially}
with a wide range of $l$. To gain intuition on how large
$\beta^{(l)}$ can be, $\beta^{(8)}$ and $\beta^{(16)}$ are
summarized for all data sets in Table \ref{tab:uniformity}. It can
be observed that $\beta^{(8)}$ and $\beta^{(16)}$ have large values,
indicating the significant unevenness of collected distributions.

\begin{table}[htb]
\begin{center}
\caption{$\beta^{(8)}$ and $\beta^{(16)}$ of collected
distributions.} \label{tab:uniformity}
\begin{tabular}{|r|c|c|c|c|c|} \hline
   $\beta^{(l)}$  & D1   & P1   & C1    & W1    & W2       \\ \hline
   $\beta^{(8)}$  & 7.9  & 8.4  & 9.0   & 12.0  & 7.8      \\
   $\beta^{(16)}$ & 31.2 & 43.2 & 52.2  & 126.7 & 50.2     \\ \hline
   \hline
   $\beta^{(l)}$  & D1-80   & D1-135   & D1-445    & D1-1023    & D1-6129       \\ \hline
   $\beta^{(8)}$  & 7.9     & 15.4     & 10.5      & 48.2       & 9.1      \\
   $\beta^{(16)}$ & 153.3   & 186.6    & 71.7      & 416.3      & 128.9     \\ \hline
\end{tabular}
\end{center}
\end{table}

\subsection{Shannon Entropy}

To further understand the importance of the non-uniformity factor,
we now turn our attention on the Shannon entropy for comparison. It
is well-known that the Shannon entropy can be used to measure the
non-uniformity of a probability distribution \cite{info}. The
Shannon entropy in /$l$ subnets is defined as
\begin{equation}
  H\left(P^{(l)}\right)=-\sum_{i=1}^{2^l}{p_g^{(l)}(i)\log_2p_g^{(l)}(i)},
\end{equation}
where $P^{(l)}=\{p_g^{(l)}(1), p_g^{(l)}(2), \cdots,
p_g^{(l)}(2^l)\}$.

It is noted that
\begin{equation}
  0 \le H\left(P^{(l)}\right) \le l.
\end{equation}
If a distribution is uniform, $H\left(P^{(l)}\right)$ achieves the
maximum value $l$. On the other hand, if a distribution is extremely
non-uniform, {\em e.g.}, all vulnerable hosts concentrate on one
subnet, $H\left(P^{(l)}\right)$ obtains the minimum value 0.

Furthermore, we compare $H\left(P^{(l)}\right)$ with
$H\left(P^{(l-1)}\right)$ and find that their difference is between
0 and 1, as shown in the following theorem.
\begin{theorem}
\label{thm:shannon}
\begin{equation}
 H\left(P^{(l-1)}\right) \le H\left(P^{(l)}\right) \le H\left(P^{(l-1)}\right)+1,
\end{equation}
where $l \in \{1,\cdots,32\}$.
\end{theorem}

The proof of Theorem \ref{thm:shannon} is given in Appendix 2. If
vulnerable hosts in each group of /$(l-1)$ subnets are extremely
unevenly divided into the groups of /$l$ subnets ({\em i.e.},
$p_g^{(l)}(2i-1)=0$ or $p_g^{(l)}(2i)=0$, $\forall i \in
\{1,2,\cdots , 2^{l-1}\}$), then $H\left(P^{(l)}\right) =
H\left(P^{(l-1)}\right)$. If the division of vulnerable hosts is
equal for all groups ({\em i.e.},
$p_g^{(l)}(2i-1)=p_g^{(l)}(2i)=\frac{1}{2}p_g^{(l-1)}(i)$, $\forall
i$)), then $H\left(P^{(l)}\right) = H\left(P^{(l-1)}\right)+1$.
Excluding these two extreme cases, $H\left(P^{(l-1)}\right) <
H\left(P^{(l)}\right) < H\left(P^{(l-1)}\right)+1$. Therefore,
$H\left(P^{(l)}\right)$ is a non-decreasing function of $l$.
Moreover, the difference between $H\left(P^{(l)}\right)$ and
$H\left(P^{(l-1)}\right)$ reflects how evenly vulnerable hosts in
each /($l-1$) subnet distribute between two groups of /$l$ subnets.

\begin{figure}[bt!]
\centering \subfigure[Five data sets.] { \label{fig:shannon_all}
\psfig{figure=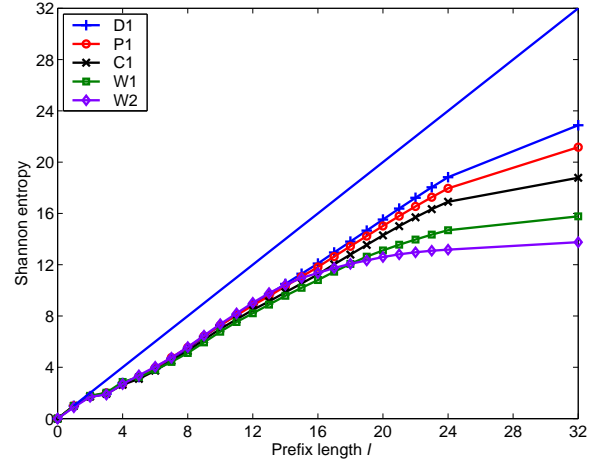,width=7.8cm} } \subfigure[D1 with
different ports.] { \label{fig:shonnon_dshield}
\psfig{figure=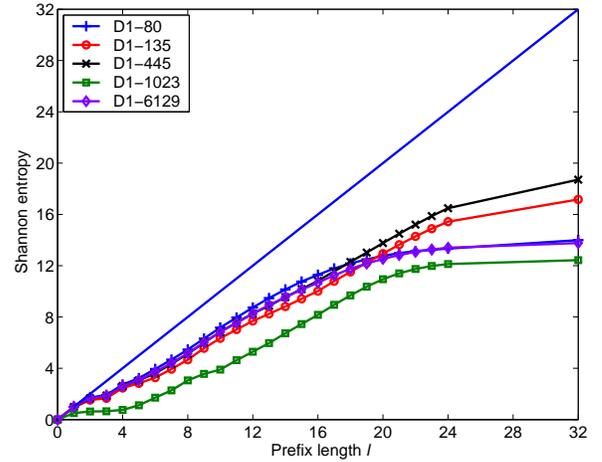,width=7.8cm} }
\caption{Shannon entropies of collected data sets.}
\label{fig:shannon}
\end{figure}

Figure \ref{fig:shannon} shows the Shannon entropies of our
empirical distributions from the data sets.
$H\left(P^{(l)}\right)=l$ is denoted by the diagonal line in the
figure. It can be seen that the curves for our collected data sets
are similar.

\subsection{Non-Uniformity Factor, Renyi Entropy, and Shannon Entropy}

To quantify the difference between the non-uniformity factor and the
Shannon entropy, we note that the non-uniformity factor directly
relates to the Renyi entropies of order two and zero, as shown in
the following equation:
\begin{equation}
\label{equ:renyi2}
  \beta^{(l)} = 2^{l-H_2\left(P^{(l)}\right)} = 2^{H_0\left(P^{(l)}\right)-H_2\left(P^{(l)}\right)},
\end{equation}
where $P^{(l)}=\{p_g^{(l)}(1),p_g^{(l)}(2),\cdots,p_g^{(l)}(2^l)
\}$. Therefore, the non-uniformity factor is essentially a Renyi
entropy. Hence, the non-uniformity factor corresponds to a
generalized entropy of order 2, whereas the Shannon entropy is the
generalized entropy of order 1.

Why do we choose the non-uniformity factor rather than the Shannon
entropy? We compare these two metrics in terms of characterizing a
vulnerable-host distribution and find the following fundamental
differences.

\begin{itemize}
\item
In Figure \ref{fig:rate}, when a distribution is uniform,
$\beta^{(l)}=1$. Hence, the distance between $\beta^{(l)}$ and the
horizontal access $1$ measures the degree of unevenness. Similarly,
the distance between $H\left(P^{(l)}\right)$ and 0 in Figure
\ref{fig:shannon} reflects how uniform a distribution is. A larger
$H\left(P^{(l)}\right)$ corresponds to a more even distribution,
whereas a larger $\beta^{(l)}$ corresponds to a more non-uniform
distribution. In addition, if two distributions have different
prefix lengths, we can directly apply the non-uniformity factor (or
the Shannon entropy) to compare the unevenness (or evenness) between
them. Therefore, the Shanon entropy provides a better measure for
describing the evenness of a distribution, while the non-uniformity
factor gives a better metric for characterizing the non-uniformity
of a distribution.
\item
From Theorem \ref{thm:comp} and Equation (\ref{equ:renyi2}), we have
$\beta^{(l)}> 2^{l-H\left(P^{(l)} \right)}$ when the non-zero
$p_g^{(l)}$'s are not all equal. Meanwhile, evidenced by Figures
\ref{fig:rate} and \ref{fig:shannon}, the non-uniformity factor
magnifies the unevenness of a distribution. Therefore, $\beta^{(l)}$
depends more on the large $p_g^{(l)}$'s.
\end{itemize}

A more important aspect of using the non-uniformity factor is its
relation to some real randomized epidemic algorithms ({\em e.g.},
localized scanning and sequential scanning). Such a relation cannot
be drawn using the Shannon entropy. We will show this in the next
section.

\section{Network-Aware Malware Spreading Ability}
\label{sec:ability}

In this section, we explicitly relate the speed of malware
propagation with the information bits extracted by random-scanning
and network-aware malwares.

\subsection{Collision Probability, Uncertainty, and Information Bits}

We begin with defining three important quantities: the collision
probability, uncertainty, and information bits. Consider a randomly
chosen vulnerable host $Y$. The probability that this host is in the
$/l$ subnet $i$ is $p_g^{(l)}(i)$. Imagine that a malware guesses
which subnet host $Y$ belongs to and chooses a target $/l$ subnet
$i$ with the probability $q_g^{(l)}(i)=p_g^{(l)}(i)$. Thus, the
probability for the malware to make a correct guess is
$p_h=\sum_{i=1}^{2^l}{\left(p_g^{(l)}(i)\right)^2}$. This
probability is called the {\it collision probability} and is defined
in \cite{Cachin}. Such a probability of success is reflected in our
designed non-uniformity factor and corresponds to a scenario that
the malware knows the underlying vulnerable-host group distribution.
Intuitively, the more non-uniform a vulnerable-host distribution is,
the larger the probability of success is, {\em i.e.}, the easier it
is for a scan to hit a vulnerable host, the more vulnerable the
network is, and the less uncertainty there is in a vulnerable-host
distribution.

We now extend the concept of the collision probability and define
$p_h$ as the probability that a malware scan hits a subnet where a
randomly chosen vulnerable host locates, {\em i.e.},
\begin{equation}
\label{equ:ph}
 p_h=\sum_{i=1}^{2^l}{p_g^{(l)}(i) q_g^{(l)}(i)}.
\end{equation}
Then two important quantities can be defined:
\begin{itemize}
\item
$-\log_2{p_h}$ as the {\em uncertainty} exhibited by the
vulnerable-host distribution $p_g^{(l)}(i)$'s.
\item
$H_0\left(P^{(l)}\right)-[-\log_2{p_h}]$ as the number of {\em
information bits} extracted by a randomized epidemic scanning
algorithm using $q_g^{(l)}(i)$'s.
\end{itemize}

Here $-\log_2{p_h}$ is regarded as the uncertainty on the
vulnerable-host distribution in view of the malware, similar to
self-information \cite{self-info}. For example, if a malware has no
information about a vulnerable-host distribution and has to use
random scanning, it has the largest uncertainty
$H_0\left(P^{(l)}\right)=l$ and extracts zero information bit from
the distribution. Likewise, the number of information bits extracted
by a network-aware malware can be measured as the reduction of the
uncertainty and thus equals to
$H_0\left(P^{(l)}\right)-[-\log_2{p_h}]$. For example,
$\log_2{\beta^{(l)}}=H_0\left(P^{(l)}\right)-H_2\left(P^{(l)}\right)$
is the information bits extractable by an adversary that chooses $
q_g^{(l)}(i)= p_g^{(l)}(i)$.

\subsection{Infection Rate}

We characterize the spread of a network-aware malware at an early
stage by deriving the infection rate. The infection rate, denoted by
$\alpha$, is defined as the average number of vulnerable hosts that
can be infected per unit time by one infected host during the early
stage of malware propagation \cite{Zou}. The infection rate is an
important metric for studying network-aware malware spreading
ability for two reasons. First, since the number of infected hosts
increases exponentially with the rate $1+\alpha$ during the early
stage, a malware with a higher infection rate can spread much faster
at the beginning and thus infect a large number of hosts in a
shorter time \cite{IS1}. Second, while it is generally difficult to
derive a close-form solution for dynamic malware propagation, we can
obtain a close-form expression of the infection rate for different
malware scanning methods.

Let $R$ denote the (random) number of vulnerable hosts that can be
infected per unit time by one infected host during the early stage
of malware propagation. The infection rate is the expected value of
$R$, {\em i.e.}, $\alpha=E[R]$. Let $s$ be the scanning rate or the
number of scans sent by an infected host per unit time, $N$ be the
number of vulnerable hosts, and $\Omega$ be the scanning space ({\em
i.e.}, $\Omega=2^{32}$).

\subsection{Random Scanning}

Random scanning (RS) has been used by most real worms. For RS, an
infected host sends out $s$ random scans per unit time, and the
probability that one scan hits a vulnerable host is
$\frac{N}{\Omega}$. Thus, $R$ follows a Binomial distribution B($s$,
$\frac{N}{\Omega}$)\footnote{In our derivation, we ignore the
dependency of the events that different scans hit the same target at
the early stage of malware propagation.}, resulting in
\begin{equation}
\label{equ:RS}
  \alpha_{RS}=E[R]=\frac{sN}{\Omega}.
\end{equation}

Another way to derive the infection rate of RS is as follows.
Consider a randomly chosen vulnerable host $Y$. The probability that
this host is in the /$l$ subnet $i$ is $p_g^{(l)}(i)$. An RS malware
can make a successful guess on which subnet host $Y$ belongs to with
collision probability
$p_h=\frac{1}{2^l}=2^{-H_0\left(P^{(l)}\right)}$. A scan from the RS
malware can be regarded as first selecting the /$l$ subnet randomly
and then choosing the host in the subnet at random. Hence the
probability for the malware to hit host $Y$ is
$\frac{1}{2^{32-l}}\cdot
2^{-H_0\left(P^{(l)}\right)}=2^{-H_0\left(P^{(l)}\right)-(32-l)}$.
Since there are $N$ vulnerable hosts, the probability for a malware
to hit a vulnerable host is $N\cdot
2^{-H_0\left(P^{(l)}\right)-(32-l)}$. Thus, $R$ follows a Binomial
distribution B($s$, $N\cdot 2^{-H_0\left(P^{(l)}\right)-(32-l)}$),
resulting in
\begin{equation}
\label{equ:RS2}
  \alpha_{RS}=E[R]=\frac{sN}{2^{32-l}}\cdot 2^{-H_0\left(P^{(l)}\right)}.
\end{equation}
Therefore, for the RS malware, the uncertainty on the
vulnerable-host distribution is
$-\log_2{p_h}=H_0\left(P^{(l)}\right)$, {\em i.e.}, the number of
information bits on vulnerable hosts extracted by RS is
$H_0\left(P^{(l)}\right)-H_0\left(P^{(l)}\right)= 0$.

\subsection{Optimal Importance Scanning}

Importance scanning (IS) exploits the non-uniform distribution of
vulnerable hosts. We derive the infection rate of IS. An infected
host scans /$l$ subnet $i$ with the probability $q_g^{(l)}(i)$.
Consider a randomly chosen vulnerable host $Y$. The probability for
this host being in /$l$ subnet $i$ is $p_g^{(l)}(i)$. An IS malware
can make a successful guess on which subnet host $Y$ belongs to with
collision probability
$p_h=\sum_{i=1}^{2^l}{p_g^{(l)}(i)q_g^{(l)}(i)}$. Thus, the
probability for the malware to hit the host $Y$ is
$\frac{1}{2^{32-l}}\sum_{i=1}^{2^l}{p_g^{(l)}(i)q_g^{(l)}(i)}$.
Similar to RS, $R$ of IS follows a Binomial distribution
B($s$,$\frac{N}{2^{32-l}}\sum_{i=1}^{2^l}{p_g^{(l)}(i)q_g^{(l)}(i)}$),
which leads to\footnote{The same result was derived in \cite{IS1}
but by a different approach.}
\begin{equation}
\label{equ:IS}
  \alpha_{IS}=E[R]=\frac{sN}{2^{32-l}}\sum_{i=1}^{2^l}{p_g^{(l)}(i)q_g^{(l)}(i)}.
\end{equation}
Therefore, the uncertainty of the vulnerable-host distribution for
an IS malware is
$-\log_2{\sum_{i=1}^{2^l}{p_g^{(l)}(i)q_g^{(l)}(i)}}$, and the
number of information bits on vulnerable hosts extracted by IS is
$H_0\left(P^{(l)}\right)+\log_2{\sum_{i=1}^{2^l}{p_g^{(l)}(i)q_g^{(l)}(i)}}$.

Note that importance scanning can choose $q_g^{(l)}(i)$'s to
maximize the infection rate, resulting in a ``worst-case" scenario
for defenders or /$l$ optimal IS (/$l$ OPT\_IS) for attackers
\cite{IS1}, {\em i.e.},
\begin{equation}
\label{equ:OPT_IS}
  \alpha_{OPT\_IS}^{(l)} =
  \max\{\alpha_{IS}\}=\frac{sN}{2^{32-l}}\max_i\{p_g^{(l)}(i)\}.
\end{equation}
That is,
\begin{equation}
  \alpha_{OPT\_IS}^{(l)} =
  \frac{sN}{2^{32-l}}2^{-H_\infty\left(P^{(l)}\right)}=\alpha_{RS}\cdot 2^{H_0\left(P^{(l)}\right)-H_\infty\left(P^{(l)}\right)}.
\end{equation}
Therefore, the uncertainty on the vulnerable-host distribution for
/$l$ OPT\_IS is $H_\infty\left(P^{(l)}\right)$; and the number of
infection bits on vulnerable hosts extracted by this scanning method
is $H_0\left(P^{(l)}\right)-H_\infty\left(P^{(l)}\right)$.

\subsection{Suboptimal Importance Scanning}

As shown in our prior work \cite{IS1}, the optimal IS is difficult
to implement in reality. Hence we also consider a special case of
IS, where the group scanning distribution $q_g^{(l)}(i)$ is chosen
to be proportional to the number of vulnerable hosts in group $i$,
{\em i.e.}, $q_g^{(l)}(i)=p_g^{(l)}(i)$. This results in suboptimal
IS \cite{IS1}, called /$l$ IS. Thus, the infection rate is
\begin{equation}
\label{equ:suboptimal_IS}
  \alpha_{IS}^{(l)} = \frac{sN}{2^{32-l}}{\sum_{i=1}^{2^l}\left(p_g(i)
  \right)^2} = \frac{sN}{2^{32-l}}{\cdot 2^{-H_2\left(P^{(l)}\right)}} = \alpha_{RS}\cdot \beta^{(l)}.
\end{equation}
Therefore, the uncertainty on the vulnerable-host distribution for
/$l$ IS is $H_2\left(P^{(l)}\right)$; and the corresponding number
of information bits extracted is
$H_0\left(P^{(l)}\right)-H_2\left(P^{(l)}\right)$ or
$\log_2{\beta^{(l)}}$. Moreover, compared with RS, this /$l$ IS can
increase the infection rate by a factor of $\beta^{(l)}$. On the
other hand, RS can be regarded as a special case of suboptimal IS
when $l=0$.

\subsection{Localized Scanning}

Localized scanning (LS) has been used by such real worms as Code Red
II and Nimda. LS is a simpler randomized algorithm that utilizes
only a few parameters rather than an underlying vulnerable-host
group distribution. We first consider a simplified version of LS,
called /$l$ LS, which scans the Internet as follows:
\begin{itemize}
\item
$p_a$ ($0\le p_a \le 1$) of the time, an address with the same first
$l$ bits is chosen as the target,
\item
$1-p_a$ of the time, an address is chosen randomly from an entire IP
address space.
\end{itemize}
Hence LS is an oblivious yet local randomized algorithm where the
locality is characterized by parameter $p_a$. Assume that an
initially infected host is randomly chosen from the vulnerable
hosts. Let $I_g$ denote the subnet where an initially infected host
locates. Thus, $P(I_g=i)=p_g^{(l)}(i)$, where $i=1,2,\cdots,2^l$.
For an infected host located in /$l$ subnet $i$, a scan from this
host probes globally with the probability of $1-p_a$ and hits /$l$
subnet $j$ ($j\not= i$) with the likelihood of $\frac{1-p_a}{2^l}$.
Thus, the group scanning distribution for this host is
\begin{equation}
\label{equ:8scan}
 q_g^{(l)}(j) = \left\{ \begin{array}{ll}
         p_a+\frac{1-p_a}{2^l}, & \mbox{if $j=i$}; \\
         \frac{1-p_a}{2^l}, & \mbox{otherwise}, \end{array} \right.
\end{equation}
where $j=1,2,\cdots,2^l$. Given the subnet location of an initially
infected host ({\em i.e.}, /$l$ subnet $i$), the {\em conditional}
collision probability or the probability for a malware scan to hit a
randomly chosen vulnerable host can be calculated based on Equation
(\ref{equ:ph}), {\em i.e.},
\begin{equation}
  p_h(i) = p_ap_g^{(l)}(i)+\frac{1-p_a}{2^l}.
\end{equation}
Therefore, we can compute the collision probability as
\begin{equation}
  p_h = \sum_{i=1}^{2^l}{P(I_g=i)p_h(i)}=p_a\sum_{i=1}^{2^l}{p_g^2(i)}+\frac{1-p_a}{2^l},
\end{equation}
resulting in
\begin{equation}
\label{equ:LS}
  \alpha_{LS}^{(l)} = \alpha_{RS} \left(1-p_a+p_a\beta^{(l)}\right).
\end{equation}
Therefore, the number of information bits extracted from the
vulnerable-host distribution by /$l$ LS is
$\log_2\{1-p_a+p_a\beta^{(l)}\}$, which is between 0 and
$H_0\left(P^{(l)}\right)-H_2\left(P^{(l)}\right)$.

Moreover, since $\beta^{(l)}
> 1$ ($\beta^{(l)}=1$ is for a uniform distribution and is excluded
here), $\alpha_{LS}^{(l)}$ increases with respect to $p_a$.
Specifically, when $p_a\rightarrow 1$, $\alpha_{LS}^{(l)}\rightarrow
\alpha_{RS}\beta^{(l)}=\alpha_{IS}^{(l)}$. Thus, /$l$ LS has an
infection rate comparable to that of /$l$ IS. In reality, $p_a$
cannot be 1. This is because an LS malware begins spreading from one
infected host that is specifically in a subnet; and if $p_a=1$, the
malware can never spread out of this subnet. Therefore, we expect
that the optimal value of $p_a$ should be large but not 1.

Next, we further consider another LS, called two-level LS (2LLS),
which has been used by the Code Red II and Nimda worms
\cite{CERT_codered,CERT_nimda}. 2LLS scans the Internet as follows:
\begin{itemize}
\item
$p_b$ ($0\le p_b \le 1$) of the time, an address with the same first
byte is chosen as the target,
\item
$p_c$ ($0\le p_c \le 1-p_b$) of the time, an address with the same
first two bytes is chosen as the target,
\item
$1-p_b-p_c$ of the time, a random address is chosen.
\end{itemize}
For example, for the Code Red II worm, $p_b = 0.5$ and $p_c = 0.375$
\cite{CERT_codered}; for the Nimda worm, $p_b = 0.25$ and $p_c =
0.5$ \cite{CERT_nimda}. Using the similar analysis for /$l$ LS, we
can derive the infection rate of 2LLS:
\begin{equation}
\label{equ:16infection}
  \alpha_{2LLS} =
  \alpha_{RS}\left(1-p_b-p_c+p_b\beta^{(8)}+p_c\beta^{(16)}\right).
\end{equation}
Similarly, the number of information bits extracted from the
vulnerable-host distribution by the 2LLS malware is
$\log_2\{1-p_b-p_c+p_b\beta^{(8)}+p_c\beta^{(16)}\}$, which is
between 0 and $H_0\left(P^{(16)}\right)-H_2\left(P^{(16)}\right)$.

Since $\beta^{(16)} \ge \beta^{(8)} \ge 1$ from Theorem
\ref{thm:uniformity}, $\alpha_{2LLS}$ holds or increases when both
$p_b$ and $p_c$ increase. Specially, when $p_c\rightarrow 1$,
$\alpha_{2LLS}\rightarrow
\alpha_{RS}\beta^{(16)}=\alpha_{IS}^{(16)}$. Thus, 2LLS has an
infection rate comparable to that of /16 IS. Moreover,
$\beta^{(16)}$ is much larger than $\beta^{(8)}$ as shown in Table
\ref{tab:uniformity} for the collected distributions. Hence, $p_c$
is more significant than $p_b$ for 2LLS.

\subsection{Modified Sequential Scanning}

The Blaster worm is a real malware that exploits sequential scanning
in combination with localized scanning. A {\it sequential-scanning}
malware studied in \cite{Zou:scan,Gu} begins to scan addresses
sequentially from a randomly chosen starting IP address and has a
similar propagation speed as a random-scanning malware. The Blaster
worm selects its starting point locally as the first address of its
Class C subnet with probability 0.4 \cite{Blaster,Zou:scan}. To
analyze the effect of sequential scanning, we do not incorporate
localized scanning. Specifically, we consider our /$l$ modified
sequential-scanning (MSS) malware, which scans the Internet as
follows:
\begin{itemize}
\item
Newly infected host $A$ begins with random scanning until finding a
vulnerable host with address $B$.
\item
After infecting the target $B$, host $A$ continues to sequentially
scan IP addresses $B+1$, $B+2$, $\cdots$ (or $B-1$, $B-2$, $\cdots$)
in the /$l$ subnet where $B$ locates.
\end{itemize}
Such a sequential malware-scanning strategy is in a similar spirit
to the {\it nearest neighbor rule}, which is widely used in pattern
classification \cite{Cover}. The basic idea is that if the
vulnerable hosts are clustered, the neighbor of a vulnerable host is
likely to be vulnerable also.

Such a /$l$ MSS malware has two stages. In the first stage (called
MSS\_1), the malware uses random scanning and has an infection rate
of $\alpha_{RS}$, {\em i.e.}, $\alpha_{MSS\_1}=\alpha_{RS}$. In the
second stage (called MSS\_2), the malware scans sequentially in a
/$l$ subnet. The fist $l$ bits of a target address are fixed,
whereas the last $32-l$ bits of the address are generated additively
or subtractively and are modulated by $2^{32-l}$. Let $I_g$ denote
the sunbet where $B$ locates. Thus, $P(I_g=i)=p_g^{(l)}(i)$, where
$i=1,2,\cdots,2^l$. Since an MSS\_2 malware scans only the subnet
where $B$ locates, the conditional collision probability
$p_h(i)=p_g^{(l)}(i)$, which leads to
$p_h=\sum_{i=1}^{2^l}{\left(p_g^{(l)}(i)\right)^2}$. Thus, the
infection rate is
\begin{equation}
\label{ineq:MSS}
  \alpha_{MSS\_2}= \alpha_{RS}\cdot \beta^{(l)}.
\end{equation}
Therefore, the uncertainty on vulnerable hosts for /$l$ MSS is
between $H_0\left(P^{(l)}\right)$ and $H_2\left(P^{(l)}\right)$.
Moreover, the infection rate of /$l$ MSS is between $\alpha_{RS}$
and $\alpha_{RS} \beta^{(l)}$.

\subsection{Summary}

In summary, the uncertainty on the vulnerable-host distribution and
the corresponding number of information bits extracted by different
randomized epidemic algorithms depends on the Renyi entropies of
different orders, as shown in Table \ref{tab:info}. Moreover, the
number of the information bits extractable by the network-aware
malwares bridges the entropy on a vulnerable-host distribution and
the malware propagation speed, as shown in the following equation.
\begin{equation}
 \mbox{Information
 bits}=H_0\left(P^{(l)}\right)-[-\log_2{p_h}]=\log_2\left\{\frac{\alpha}{\alpha_{RS}}\right\},
\end{equation}
where $p_h$ is the collision probability and $\alpha$ is the
infection rate of the malware.
\begin{table*}[htb]
\begin{center}
\caption{Summary of Randomized Epidemic Scanning Algorithms and
Information Bits } \label{tab:info}
\begin{tabular}{|r|c|c|} \hline
   Algorithm  &  Infection Rate                                                               & Information Bits Extracted       \\ \hline
      RS      &  $\frac{sN}{2^{32-l}}\cdot 2^{-H_0\left(P^{(l)}\right)}$                      & $0$      \\
 /$l$ OPT\_IS &  $\alpha_{RS}\cdot 2^{H_0\left(P^{(l)}\right)-H_\infty\left(P^{(l)}\right)} $ & $H_0\left(P^{(l)}\right)-H_\infty\left(P^{(l)}\right)$ \\
  /$l$ IS     &  $\alpha_{RS}\cdot \beta^{(l)}$                                               & $\log_2\{\beta^{(l)}\}$ \\
  /$l$ LS     &  $\alpha_{RS} \left(1-p_a+p_a\beta^{(l)}\right)$                              & $\log_2\{1-p_a+p_a\beta^{(l)}\}$\\
  2LLS        &  $\alpha_{RS}\left(1-p_b-p_c+p_b\beta^{(8)}+p_c\beta^{(16)}\right)$           & $\log_2\{1-p_b-p_c+p_b\beta^{(8)}+p_c\beta^{(16)}\}$ \\
  /$l$ MSS\_2 &  $\alpha_{RS}\cdot \beta^{(l)}$                                               & $\log_2\{\beta^{(l)}\}$ \\  \hline
\end{tabular}
\end{center}
\end{table*}

When /$l$ subnets are considered, RS has the largest uncertainty
$H_0\left(P^{(l)}\right)$, while optimal IS has the smallest
uncertainty $H_\infty\left(P^{(l)}\right)$. Moreover, infection
rates of all three network-aware malwares (suboptimal IS, LS, and
MSS) can be far larger than that of an RS malware, depending on the
non-uniformity factors ({\em i.e.}, $\beta^{(l)}$) or the Renyi
entropy in the order of two ({\em i.e.}, $H_2\left(P^{(l)}\right)$).
The infection rates of all these scanning algorithms are
characterized by the Renyi entropies, relating the efficiency of a
randomized scanning algorithm with the information bits in a
non-uniform vulnerable-host distribution.

Hence we relate the information theory with the network-aware
malware propagation through the Renyi entropy. The uncertainty of a
vulnerable-host group probability distribution, which is quantified
by the Renyi entropy, is important for an attacker to design a
network-aware malware. If there is no uncertainty about the
distribution of vulnerable hosts ({\em e.g.}, either all vulnerable
hosts are concentrated on a subnet or all information about
vulnerable hosts is known), the entropy is minimum, and the malware
that uses the information on the distribution can spread fastest by
employing the optimal importance scanning. On the other hand, if
there is maximum uncertainty ({\em e.g.}, vulnerable hosts are
uniformly distributed), the entropy is maximum. For this case, the
best a malware can take an advantage from a uniform distribution is
to use random scanning. In general, when an attacker obtains more
information about a non-uniform vulnerable-host distribution, the
resulting malware can spread faster.

\section{Simulation and Validation}
\label{sec:simulations}

We now validate our analytical results through simulations on the
collected data sets.

\subsection{Infection Rate}

We first focus on validating infection rates. We apply the discrete
event simulation to our experiments \cite{Ross}. Specifically, we
simulate the searching process of a malware using different scanning
methods at the early stage. We use the C1 data set for the
vulnerable-host distribution. Note that the C1 distribution has the
non-uniformity factors $\beta^{(8)} = 9.0$ and $\beta^{(16)}=52.2$,
and $\max_i\{p_g^{(l)}(i)\}=0.0041$. The malware spreads over the C1
distribution with $N=448,894$ and has a scanning rate $s=100$. The
uncertainty on the vulnerable-host distribution and the information
bits extractable for different scanning methods are shown in Table
\ref{tab:simulation}. The simulation stops when the malware has sent
out $10^3$ scans for RS, /16 OPT\_IS, /16 IS, /16 LS, and 2LLS, and
65,535 scans for /16 MSS\_2. Then, we count the number of vulnerable
hosts hit by the malware scans and compute the infection rate. The
results are averaged over $10^4$ runs. Table \ref{tab:simulation}
compares the simulation results ({\em i.e.}, sample mean) with the
analytical results ({\em i.e.}, Equations (\ref{equ:RS2}),
(\ref{equ:OPT_IS}), (\ref{equ:suboptimal_IS}), (\ref{equ:LS}),
(\ref{equ:16infection}), and (\ref{ineq:MSS})). Here, a /16 LS
malware uses $p_a=0.75$, whereas a 2LLS malware employs $p_b=0.25$
and $p_c=0.5$. We observe that the sample means and the analytical
results are almost identical.
\begin{table*}[htb]
\begin{center}
\caption{Uncertainty on the vulnerable-host distribution,
information bits, and infection rates of different scanning
methods.} \label{tab:simulation}
\begin{tabular}{|r|c|c|c|c|c|c|} \hline
   Scanning method                      & RS      & /16 OPT\_IS & /16 IS  & /16 LS  & 2LLS    & /16 MSS\_2 \\ \hline
   Uncertainty (analytical result)      & 16      &   7.9266    & 10.2940 & 10.6999 & 11.1620 & 10.2940  \\
   Information bits (analytical result) & 0       &   8.0734    & 5.7060  & 5.3001  & 4.8380  & 5.7060 \\
   Infection rate (analytical result)   & 0.0105  &   2.8152    & 0.5456  & 0.4118  & 0.2989  & 0.5456      \\
   Infection rate (sample mean)         & 0.0103  &   2.7745    & 0.5454  & 0.4023  & 0.2942  & 0.5489     \\
   Infection rate (sample variance)     & 0.0010  &   0.2597    & 0.0543  & 0.2072  & 0.1053  & 0.3186     \\ \hline
\end{tabular}
\end{center}
\end{table*}

We observe that network-aware malwares have much larger infection
rates than random-scanning malwares. LS indeed increases the
infection rate with nearly a non-uniformity factor and approaches
the capacity of suboptimal IS. This is significant as LS only
depends on one or two parameters ({\em i.e.}, $p_a$ for /$l$ LS and
$p_b$, $p_c$ for 2LLS), while IS requires the information of the
vulnerable-host distribution. On the other hand, LS has a larger
sample variance than IS as indicated by Table \ref{tab:simulation}.
This implies that the infection speed of an LS malware depends on
the location of initially infected hosts. If the LS malware begins
spreading from a subnet containing densely populated vulnerable
hosts, the malware would spread rapidly. Furthermore, we notice that
the MSS malware also has a large infection rate at the second stage,
indicating that MSS can indeed exploit the clustering pattern of the
distribution. Meanwhile, the large sample variance of the infection
rate of MSS\_2 reflects that an MSS malware strongly depends on the
initially infected hosts. We further compute the infection rate of a
/16 MSS malware that includes both random-scanning and
sequential-scanning stages. Simulation results are averaged over
$10^6$ runs and are summarized in Table \ref{tab:MSS}. These results
strongly depend on the total number of malware scans. When the
number of malware scans is small, an MSS malware behaves similar to
a random-scanning malware. When the number of malware scans
increases, the MSS malware spends more scans on the second stage and
thus has a larger infection rate.
\begin{table}[htb]
\begin{center}
\caption{Infection rates of a /16 MSS malware.} \label{tab:MSS}
\begin{tabular}{|r|c|c|c|c|c|} \hline
   \# of malware scans     & 10     & 100    & 1000    & 10000   & 50000       \\ \hline
   Sample mean         & 0.0108  & 0.0190 & 0.0728  & 0.2866  & 0.4298     \\
   Sample variance     & 0.1246  & 0.1346 & 0.1659  & 0.2498  & 0.2311     \\ \hline
\end{tabular}
\end{center}
\end{table}

\subsection{Dynamic Malware Propagation}

An infection rate only characterizes the early stage of malware
propagation. We now employ the analytical active worm propagation
(AAWP) model and its extensions to characterize the entire spreading
process of malwares \cite{AAWP}. Specifically, the spread of RS and
IS malwares is implemented as described in \cite{IS1}, whereas the
propagation of LS malwares is modeled according to \cite{Rajab}. The
parameters that we use to simulate a malware are comparable to those
of the Code Red v2 worm. Code Red v2 has a vulnerable population
$N=360,000$ and a scanning rate $s=358$ per minute
\cite{Zou:codered}. We assume that the malware begins spreading from
an initially infected host that is located in the subnet containing
the largest number of vulnerable hosts.

\begin{figure}[tb]
\centerline{\psfig{figure=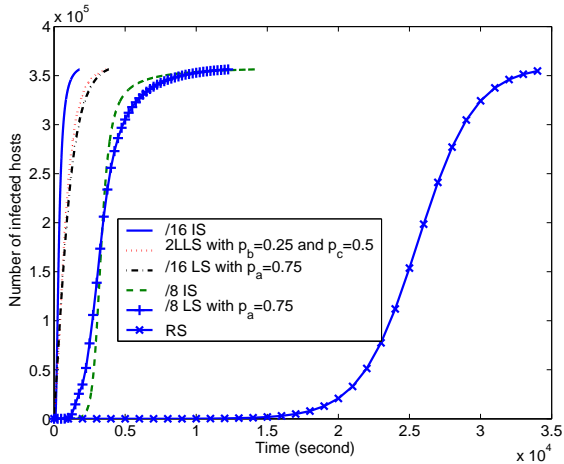,width=7.5cm}} \caption{A
network-aware malware spreads over the D1-80 distribution.}
\label{fig:methods}
\end{figure}

We first show the propagation speeds of network-aware malwares for
the same vulnerable-host distribution from data set D1-80. From
Section \ref{sec:ability}, we expect that a network-aware malware
can spread much faster than an RS malware. Figure \ref{fig:methods}
demonstrates such an example on a malware that uses different
scanning methods. It takes an RS malware 10 hours to infect 99\% of
vulnerable hosts, whereas a /8 LS malware with $p_a=0.75$ or a /8 IS
malware takes only about 3.5 hours. A /16 LS malware with $p_a=0.75$
or a 2LLS malware with $p_b=0.25$ and $p_c=0.5$ can further reduce
the time to 1 hour. A /16 IS malware spreads fastest and takes only
0.5 hour.

\begin{figure}[tb]
\centerline{\psfig{figure=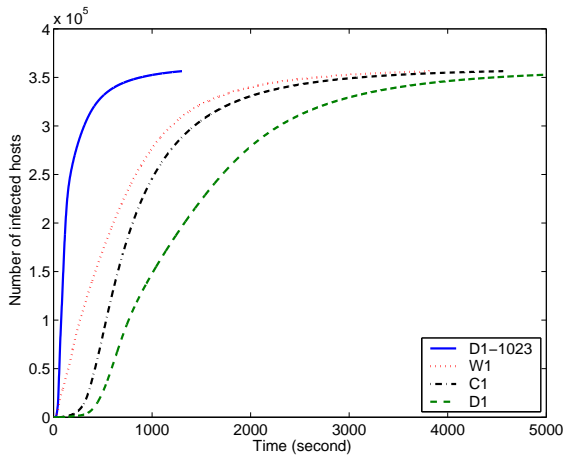,width=7.5cm}} \caption{A
2LLS malware spreads over different distributions.}
\label{fig:distribution}
\end{figure}

We also study the effect of vulnerable-host distributions on the
propagation of network-aware malwares. From Table
\ref{tab:uniformity}, we observe that
$\beta_{D1-1023}^{(16)}>\beta_{W1}^{(16)}>\beta_{C1}^{(16)}>\beta_{D1}^{(16)}$.
Thus, we expect that a network-aware malware using the /16 D1-1023
distribution would spread faster than using other three
distributions. Figure \ref{fig:distribution} verifies this through
the simulations of the spread of a 2LLS malware that uses different
vulnerable-host distributions ({\em i.e.}, D1-1023, W1, C1, and D1).
Here, the 2LLS malware employs the same parameters as the Nimda
worm, {\em i.e.}, $p_b=0.25$ and $p_c=0.5$. As expected, the malware
using the D1-1023 distribution spreads fastest, especially at the
early stage of malware propagation.

\section{Effectiveness of defense strategies}
\label{sec:applications}

What are new requirements and challenges for a defense system to
slow down the spread of a network-aware malware? We study the
effectiveness of defense strategies through non-uniformity factors.

\subsection{Host-Based Defense}

Host-based defense has been widely used for random-scanning
malwares. Proactive protection and virus throttling are examples of
host-based defense strategies.

A {\it proactive protection} (PP) strategy proactively hardens a
system, making it difficult for a malware to exploit vulnerabilities
\cite{Brumley}. Techniques used by PP include address-space
randomization, pointer encryption, instruction-set randomization,
and password protection. Thus, a malware requires multiple trials to
compromise a host that implements PP. Specifically, let $p$ ($0\le p
\le 1$) denote the protection probability or the probability that a
single malware attempt succeeds in infecting a vulnerable host that
implements PP. On the average, a malware should make $\frac{1}{p}$
exploit attempts to compromise the target. We assume that hosts with
PP are uniformly deployed in the Internet. Let $d$ ($0 < d \le 1$)
denote the deployment ratio of the number of hosts with PP to the
total number of hosts.

To show the effectiveness of the PP strategy, we consider the
infection rate of a /$l$ IS malware. Since now some of the
vulnerable hosts implement PP, Equation (\ref{equ:suboptimal_IS})
changes to
\begin{eqnarray}
\nonumber
  \alpha_{IS}^{(l)} & = & \frac{sN}{2^{32-l}}\sum_{i=1}^{2^l}{\left[ d p \left(p_g^{(l)}(i) \right) ^2 + (1-d)\left(p_g^{(l)}(i) \right) ^2 \right]} \\
\label{equ:defense}
                    & = & \alpha_{RS}\beta^{(l)}(1-d+dp).
\end{eqnarray}
To slow down the spread of a suboptimal IS malware to that of a
random-scanning malware, $\beta^{(l)}(1-d+dp) \le 1$, resulting in
\begin{equation}
\label{equ:require}
  p \le \frac{1-(1-d)\beta^{(l)}}{d\beta^{(l)}}.
\end{equation}
When PP is fully deployed, {\em i.e.}, $d=1$, $p$ can be at most
$\frac{1}{\beta^{(l)}}$. On the other hand, if PP provides perfect
protection, {\em i.e.}, $p=0$, $d$ should be at least
$1-\frac{1}{\beta^{(l)}}$. Therefore, when $\beta^{(l)}$ is large,
Inequality (\ref{equ:require}) presents high requirements for the PP
strategy. For example, if $\beta^{(16)}=50$ (most of
$\beta^{(16)}$'s in Table \ref{tab:uniformity} are larger than this
value), $p \le 0.02$ and $d \ge 0.98$. That is, a PP strategy should
be almost fully deployed and provide a nearly perfect protection for
a vulnerable host.

\begin{figure}[tb]
\centerline{\psfig{figure=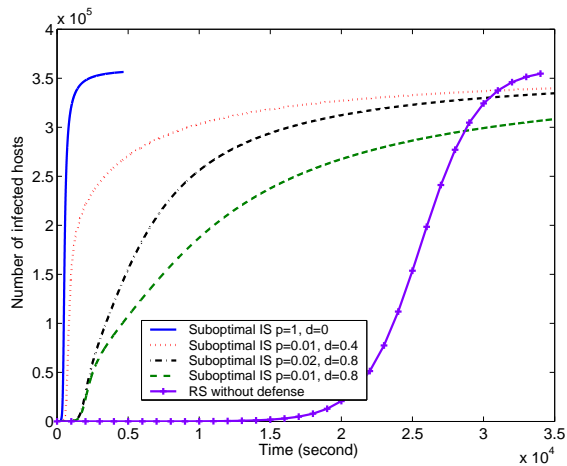,width=7.5cm}}
\caption{A /16 IS malware spreads under the defense of PP.}
\label{fig:defense}
\end{figure}

We extend the model described in \cite{IS1} to characterize the
spread of suboptimal IS malwares under the defense of the PP
strategy and show the results in Figure \ref{fig:defense}. Here,
Code-Red-v2-like malwares spread over the C1 distribution with
$\beta^{(16)}=52.2$. It is observed that even when the protection
probability is small ({\em e.g.}, $p=0.01$) and the deployment ratio
is high ({\em e.g.}, $d=0.8$), a /16 IS malware is slowed down a
little at the early stage, compared with a /16 IS malware without
the PP defense ({\em i.e.}, $p=1$ and $d=0$). Moreover, when $p$ is
small ({\em e.g.}, $p\le 0.02$), $d$ is a more sensitive parameter
than $p$.

We next consider the {\it virus throttling} (VT) strategy
\cite{Twycross}. VT constrains the number of outgoing connections of
a host and can thus reduce the scanning rate of an infected host. We
find that Equation (\ref{equ:defense}) also holds for this strategy,
except that $p$ is the ratio of the scanning rate of infected hosts
with VT to that of infected hosts without VT. Therefore, VT also
requires to be almost fully deployed for fighting network-aware
malwares effectively.

From these two strategies, we have learned that an effective
strategy should reduce either $\alpha_{RS}$ or $\beta^{(l)}$.
Host-based defense, however, is limited in such capabilities.

\subsection{IPv6}

IPv6 can decrease $\alpha_{RS}$ significantly by increasing the
scanning space \cite{Zou}. But the non-uniformity factor would
increase the infection rate if the vulnerable-host distribution is
still non-uniform. Hence, an important question is whether IPv6 can
counteract network-aware malwares when both $\alpha_{RS}$ and
$\beta^{(l)}$ are taken into consideration.

We study this issue by computing the infection rate of a
network-aware malware in the IPv6 Internet. As pointed out by
\cite{Bellovin}, a smart malware can first detect some vulnerable
hosts in /64 subnets containing many vulnerable hosts, then release
to the hosts on the hitlist, and finally spread inside these
subnets. Such a malware only scans the local /64 subnet. Thus, we
focus on the spreading speed of a network-aware malware in a /64
subnet. From Figure \ref{fig:rate}, we extrapolate that
$\beta^{(32)}$ in the IPv6 Internet can be in the order of $10^{5}$
if hosts are still distributed in a clustered fashion. Using the
parameters $N=10^8$ proposed by \cite{Feng} and $s=4,000$ used by
the Slammer worm \cite{Moore:03}, we derive the infection rate of a
/32 IS malware in a /64 subnet of the IPv6 Internet:
$\alpha_{IS}^{IPv6}=\frac{sN}{2^{64}}\cdot \beta^{(32)}=2.2\times
10^{-3}$. $\alpha_{IS}^{IPv6}$ is larger than the infection rate of
the Code Red v2 worm in the IPv4 Internet, where
$\alpha_{RS}^{CR}=\frac{360,000\times 358/60}{2^{32}}=5\times
10^{-4}$.

Therefore, IPv6 can only slow down the spread of a network-aware
malware to that of a random-scanning malware in IPv4. To defend
against the malware effectively, we should further consider how to
slow down the increase rate of $\beta^{(l)}$ as $l$ increases when
IPv4 is updated to IPv6. In essence, we should reduce the
information bits extractable by the network-aware malwares from the
vulnerable-host distribution.

\section{Conclusions}
\label{sec:conclusions}

In this paper, we have first obtained and characterized empirical
vulnerable-host distributions, using five large measurement sets
from different sources. We have derived a simple metric, known as
the non-uniformity factor, to quantify an uneven distribution of
vulnerable hosts. The non-uniformity factors have been obtained on
the empirical vulnerable-host distributions using our collected
data, and all of which demonstrate large values. This implies that
the non-uniformity of the vulnerable-host distribution is
significant and seems to be consistent across networks and
applications. Moreover, the non-uniformity factor, shown as a
function of the Renyi entropies of order two and zero, better
characterizes the uneven feature of a distribution than the Shannon
entropy.

We have drawn a relationship between Renyi entropies and randomized
epidemic scanning algorithms. In particular, we have quantified the
spreading ability of network-aware malwares that utilize randomized
scanning algorithms at the early stage. These randomized
malware-scanning algorithms range from optimal randomized scanning
({\em e.g.}, importance scanning) to real malware scanning ({\em
e.g.}, localized scanning). We have derived analytical expressions
relating the infection rates of network-aware malwares with the
uncertainty ({\em i.e.}, Renyi entropy) of finding vulnerable hosts.
We have derived and empirically verified that localized scanning and
modified sequential scanning can increase the infection rate by
nearly a non-uniformity factor when compared to random scanning and
thus approach the capacity of suboptimal importance scanning. As a
result, we have bridged the information bits extracted by malwares
from a vulnerable-host distribution with the propagation speed of
network-aware malwares.

Furthermore, we have evaluated the effectiveness of several commonly
used defense strategies on network-aware malwares. The host-based
defense, such as proactive protection or virus throttling, requires
to be almost fully deployed to slow down malware spreading at the
early stage. This implies that host-based defense would be weakened
significantly by network-aware scanning. More surprisingly,
different from previous findings, we have shown that network-aware
malwares can be zero-day malwares in the IPv6 Internet if vulnerable
hosts are still clustered. These findings present a significant
challenge to malware defense: Entirely different strategies may be
needed for fighting against network-aware malwares.

The information-theoretical view of malware attacks provides us a
quantification and a better understanding of three aspects: a
non-uniform vulnerable-host distribution characterized by the
non-uniformity factor or the Renyi entropy of order two, randomized
malware-scanning algorithms characterized by the infection rate or
the Renyi entropy of different orders, and the effectiveness of
defense strategies.

As part of our ongoing work, we plan to study in more depth
relationships between information theory and dynamic malware attacks
for developing effective detection and defense systems that would
take vulnerable-host distributions into consideration.

\section*{ACKNOWLEDGEMENT}
We are grateful to Paul Barford for providing us with the data sets
on vulnerable-host distributions and his insightful comments on this
work. We also thank CAIDA for making the Witty-worm data available.
This work was supported in part by NSF ECS 0300605.

\section*{APPENDIX}  
\renewcommand{\theequation}{A-\arabic{equation}}
\setcounter{equation}{0}  

\section*{APPENDIX 1: Proof of Theorem \ref{thm:uniformity}}
\label{app:uniform}

\begproof Group $i$ ($i=1,2,\cdots,2^{l-1}$) of /$(l-1)$ subnets is
partitioned into groups $2i-1$ and $2i$ of /$l$ subnets. Thus,
\begin{equation}
  p_g^{(l-1)}(i) = p_g^{(l)}(2i-1)+p_g^{(l)}(2i),
\end{equation}
where $i=1,2,\cdots,2^{i-1}$. Then, $\beta^{(l)}$ is related to
$\beta^{(l-1)}$ by the Cauchy-Schwarz inequality.
\begin{eqnarray}
\nonumber
  \beta^{(l)} & = & 2^{l-1} \sum_{i=1}^{2^{l-1}} \left\{ \left(\sum_{j=1}^2 {1^2}\right) \left[\sum_{j=1}^2 \left(p_g^{(l)}(2(i-1)+j)
              \right)^2\right]
              \right\} \\
\nonumber
              & \ge & 2^{l-1} \sum_{i=1}^{2^{l-1}}{\left(\sum_{j=1}^{2}{p_g^{(l)}(2(i-1)+j)}
              \right)^2} \\
              & = & \beta^{(l-1)}.
\end{eqnarray}
The equality holds when
$p_g^{(l)}(2i-1)=p_g^{(l)}(2i)=\frac{p_g^{(l-1)}(i)}{2}$,
$i=1,2,\cdots,2^{l-1}$. That is, in each /$(l-1)$ subnet, the
vulnerable hosts are uniformly distributed in two groups of /$l$
subnets.

On the other hand, since
\begin{equation}
  \left( p_g^{(l)}(2i-1) \right)^2+\left( p_g^{(l)}(2i) \right)^2 \le \left( p_g^{(l)}(2i-1)+p_g^{(l)}(2i) \right)^2,
\end{equation}
we have
\begin{eqnarray}
\nonumber
 \frac{\beta^{(l)}}{\beta^{(l-1)}} & = & 2\cdot \frac{\sum_{i=1}^{2^{l-1}}\left[ \left(p_g^{(l)}(2i-1)\right)^2+ \left(p_g^{(l)}(2i) \right)^2
 \right]}{\sum_{i=1}^{2^{l-1}}{\left(p_g^{(l-1)}(i)\right)^2}} \\
     & \le & 2.
\end{eqnarray}
The equality holds when for $\forall i$, $p_g^{(l)}(2i-1)=0$ or
$p_g^{(l)}(2i)=0$. That is, in each /$(l-1)$ subnet, the vulnerable
hosts are extremely non-uniformly distributed in two groups of /$l$
subnets.
\endproof

\section*{APPENDIX 2: Proof of Theorem \ref{thm:shannon}}

\begproof Since
\begin{equation}
  p_g^{(l-1)}(i) = \sum_{j=1}^{2}{p_g^{(l)}(2(i-1)+j)} \ge p_g^{(l)}(2(i-1)+j),
\end{equation}
where $i=1,2,\cdots,2^{l-1}$, we have
\begin{align}
\nonumber
  &H\left(P^{(l)}\right) \\
\nonumber
  &= -\sum_{i=1}^{2^{l-1}}{\sum_{j=1}^{2}{p_g^{(l)}(2(i-1)+j)\log_2p_g^{(l)}(2(i-1)+j)}} \\
\nonumber
  &\ge -\sum_{i=1}^{2^{l-1}}{\sum_{j=1}^{2}{p_g^{(l)}(2(i-1)+j)\log_2p_g^{(l-1)}(i)}} \\
  &= H\left(P^{(l-1)}\right).
\end{align}
The equality holds when for $\forall i$, $p_g^{(l)}(2i-1)=0$ or
$p_g^{(l)}(2i)=0$. That is, in each /$(l-1)$ subnet, the vulnerable
hosts are extremely non-uniformly distributed in two groups of /$l$
subnets.

On the other hand, using the log-sum inequality,
\begin{align}
\nonumber
  &\sum_{j=1}^{2}{p_g^{(l)}(2(i-1)+j) \log_2{p_g^{(l)}(2(i-1)+j)}} \\
  &\ge \left(\sum_{j=1}^{2}{p_g^{(l)}(2(i-1)+j)} \right) \log_2{\frac{\sum_{j=1}^{2}{p_g^{(l)}(2(i-1)+j)}}{2}},
\end{align}
we have
\begin{align}
\nonumber
  &H\left(P^{(l)}\right) \\
\nonumber
  &= -\sum_{i=1}^{2^{l-1}}{\sum_{j=1}^{2}{p_g^{(l)}(2(i-1)+j)\log_2p_g^{(l)}(2(i-1)+j)}} \\
\nonumber
  &\le -\sum_{i=1}^{2^{l-1}}{\left(p_g^{(l-1)}(i) \right) \log_2{\frac{p_g^{(l-1)}(i)}{2}}} \\
  &= H\left(P^{(l-1)}\right)+1.
\end{align}
The equality holds when for $\forall i$,
$p_g^{(l)}(2i-1)=p_g^{(l)}(2i)=\frac{p_g^{(l-1)}(i)}{2}$. That is,
in each /$(l-1)$ subnet, the vulnerable hosts are uniformly
distributed in two groups of /$l$ subnets.
\endproof

\begin{IEEEbiography}{Zesheng Chen}
is an Assistant Professor at the Department of Electrical and
Computer Engineering at Florida International University. He
received his M.S. and Ph.D. degrees from the School of Electrical
and Computer Engineering at the Georgia Institute of Technology in
2005 and 2007, under the supervision of Dr. Chuanyi Ji. He also
holds B.E. and M.E. degrees from the Department of Electronic
Engineering at Shanghai Jiao Tong University, Shanghai, China in
1998 and 2001, respectively.

His research interests include network security and the performance
evaluation of computer networks.
\end{IEEEbiography}

\begin{IEEEbiography}{Chuanyi Ji}
(S'85-M'91-SM'07) is an Associate Professor at the School of
Electrical and Computer Engineering at the Georgia Institute of
Technology. She received a BS (Honor) from Tsinghua University,
Beijing, China in 1983, an MS from the University of Pennsylvania,
Philadelphia in 1986, and a PhD from the California Institute of
Technology, Pasadena, in 1992, all in Electrical Engineering.
Chuanyi Ji was on the faculty at Rensselaer Polytechnic Institute
from 1991 to 2001. She spent her sabbatical at Bell-labs Lucent in
1999, and was a visiting faculty at MIT in Fall 2000.

Chuanyi Ji's research lies in the areas of network management and
control, and adaptive learning systems. Her research interests are
in understanding and managing complex networks, applications of
adaptive learning systems to network management, learning
algorithms, statistics, and information theory. Chuanyi Ji received
CAREER award from NSF in 1995 and Early Career award from Rensselaer
Polytechnic Institute in 2000.
\end{IEEEbiography}

\end{document}